\documentclass{article}

\usepackage{arxiv}

\usepackage[utf8]{inputenc}
\usepackage{geometry}
\usepackage{booktabs}
\usepackage{longtable}
\usepackage{graphicx}
\usepackage{subcaption}
\usepackage[style=numeric, backend=biber, sorting=none, maxcitenames=99, maxbibnames=99]{biblatex}
\usepackage{hyperref}
\usepackage{tabularx}
\graphicspath{ {./figures/} }
\usepackage{graphicx}
\usepackage{subcaption}

\usepackage[style=numeric, backend=biber, sorting=none, maxcitenames=99, maxbibnames=99]{biblatex}
\usepackage[affil-it]{authblk}

\title{A Synoptic Review of High-Frequency Oscillations as a Biomarker in Neurodegenerative Disease}

\author[1]{Samin Yaser}
\author[1]{Mahad Ali}
\author[3]{Yang Jiang}
\author[4]{Phuc Nguyen}
\author[5]{Jing Xiang}
\author[1, 2, *]{Laura J. Brattain}

\affil[1]{Department of Electrical and Computer Engineering, University of Central Florida, Orlando, FL, USA}
\affil[2]{Department of Medicine, University of Central Florida, Orlando, FL, USA}
\affil[3]{Department of Behavioral Science, University of Kentucky College of Medicine, Lexington, KY, USA}
\affil[4]{Manning College of Information and Computer Sciences, University of Massachusetts Amherst, MA, USA}
\affil[5]{Department of Pediatrics, University of Cincinnati, OH, USA}
\affil[*]{Corresponding author: Laura J. Brattain, laura.brattain@ucf.edu}

\begin{document}
\maketitle
\begin{abstract}
High Frequency Oscillations (HFOs), rapid bursts of brain activity above 80 Hz, have emerged as a highly specific biomarker for epileptogenic tissue. Recent evidence suggests that HFOs are also present in Alzheimer's Disease (AD), reflecting underlying network hyperexcitability and offering a promising, noninvasive tool for early diagnosis and disease tracking. This synoptic review provides a comprehensive analysis of publicly available electroencephalography (EEG) datasets relevant to HFO research in neurodegenerative disorders. We conducted a bibliometric analysis of 1,222 articles, revealing a significant and growing research interest in HFOs, particularly within the last ten years. We then systematically profile and compare key public datasets, evaluating their participant cohorts, data acquisition parameters, and accessibility, with a specific focus on their technical suitability for HFO analysis. Our comparative synthesis highlights critical methodological heterogeneity across datasets, particularly in sampling frequency and recording paradigms, which poses challenges for cross-study validation, but also offers opportunities for robustness testing. By consolidating disparate information, clarifying nomenclature, and providing a detailed methodological framework, this review serves as a guide for researchers aiming to leverage public data to advance the role of HFOs as a cross-disease biomarker for AD and related conditions.
\end{abstract}


\section{Introduction}

\subsection{High-Frequency Oscillations}
High-Frequency Oscillations (HFOs) are patterns of brain activity, detectable via electroencephalography (EEG), that occur at frequencies above the traditionally studied bands \cite{Zijlmans2012}. Foundational intracranial EEG studies identified these rapid, transient bursts of neuronal activity, typically categorized into two main types: ripples (80--250 Hz) and the more pathologically-linked fast ripples (250--500 Hz) \cite{Bragin1999, Staba2002, Worrell2011, Jacobs2011}. While physiological ripples are deeply involved in normal cognitive processes such as memory consolidation \cite{Buzsaki2015}, pathological HFOs appear to be a highly specific biomarker for epileptogenic brain tissue \cite{Jacobs2011, Gerstl2023, Frauscher2018}. Their detection and analysis, particularly from non-invasive scalp EEG, represent a significant frontier in clinical neurophysiology \cite{Maeda2025}. The distinction between pathological HFOs and traditional interictal spikes is crucial; the mechanisms generating spikes are thought to differ from those generating HFOs \cite{deCurtis2009, Staley2006}, and studies at the single-neuron level indicate that spikes co-occurring with HFOs have distinct firing patterns, suggesting they represent a more pathological form of neuronal activity \cite{Guth2021}.

\subsection{HFOs in Epilepsy and Alzheimer's Disease}
For decades, HFOs have been established as a reliable biomarker for localizing the epileptogenic zone in patients with refractory epilepsy, often guiding surgical planning with greater precision than traditional interictal spikes alone \cite{Worrell2011, Gerstl2023, Zijlmans2012}. Resection of brain tissue that generates high rates of HFOs is strongly associated with positive surgical outcomes \cite{Haegelen2013, Holler2015, vantKlooster2015, Zijlmans2012, Wu2010}. The core principle is that the brain tissue capable of generating pathological HFOs is the same tissue that initiates seizures. This link is so strong that the co-occurrence of interictal spikes and HFOs is considered a highly accurate predictor of post-surgical seizure freedom \cite{Gerstl2023}. Furthermore, interictal epileptiform discharges are not benign; they are themselves linked to transient and chronic cognitive impairments, making their detection and treatment a critical aspect of epilepsy care \cite{Holmes2013, Kleen2013}.

More recently, this concept has been extended to the study of Alzheimer's Disease (AD). A growing body of evidence suggests that neuronal network hyperexcitability is a core pathophysiological feature of AD, present even in the earliest, preclinical stages \cite{Vossel2016, Palop2007}. This hyperexcitability can manifest as subclinical epileptiform activity, which has been shown to accelerate cognitive decline \cite{Vossel2016, Tamilia2021}. Crucially, research in animal models of AD has demonstrated the presence of HFOs that are indistinguishable from those seen in epilepsy models, and that reducing this hyperexcitability can rescue cognitive deficits \cite{Lisgaras2023, Verret2012}. This finding suggests that HFOs are not specific to epilepsy but are a more general marker of network hyperexcitability. Therefore, detecting HFOs in individuals at risk for or in the early stages of AD leads to a powerful, non-invasive biomarker for early diagnosis, tracking disease progression, and assessing the efficacy of therapeutic interventions aimed at stabilizing neuronal networks \cite{Lisgaras2023, Vossel2016}.

\subsection{Bibliometric Trends in HFO Research}
To gauge the trajectory of research in this field, a bibliometric analysis was performed using the Web of Science database. The query 
\begin{verbatim}
TS=("High Frequency Oscillation" OR "HFO" OR "high frequency activity" 

OR "oscillatory biomarker" OR "interictal spikes")

AND TS=("EEG" OR "electroencephalography") 

AND TS=("Alzheimer's disease" OR "Dementia" 

OR "Epilepsy" OR "Cognitive Decline" 

OR "Neurodegeneration" OR "Parkinson")
\end{verbatim}
yielded 1,222 papers. The analysis reveals a clear and accelerating interest in HFOs as a biomarker.

\begin{figure}[h!]
    \centering
    \includegraphics[width=0.8\textwidth]{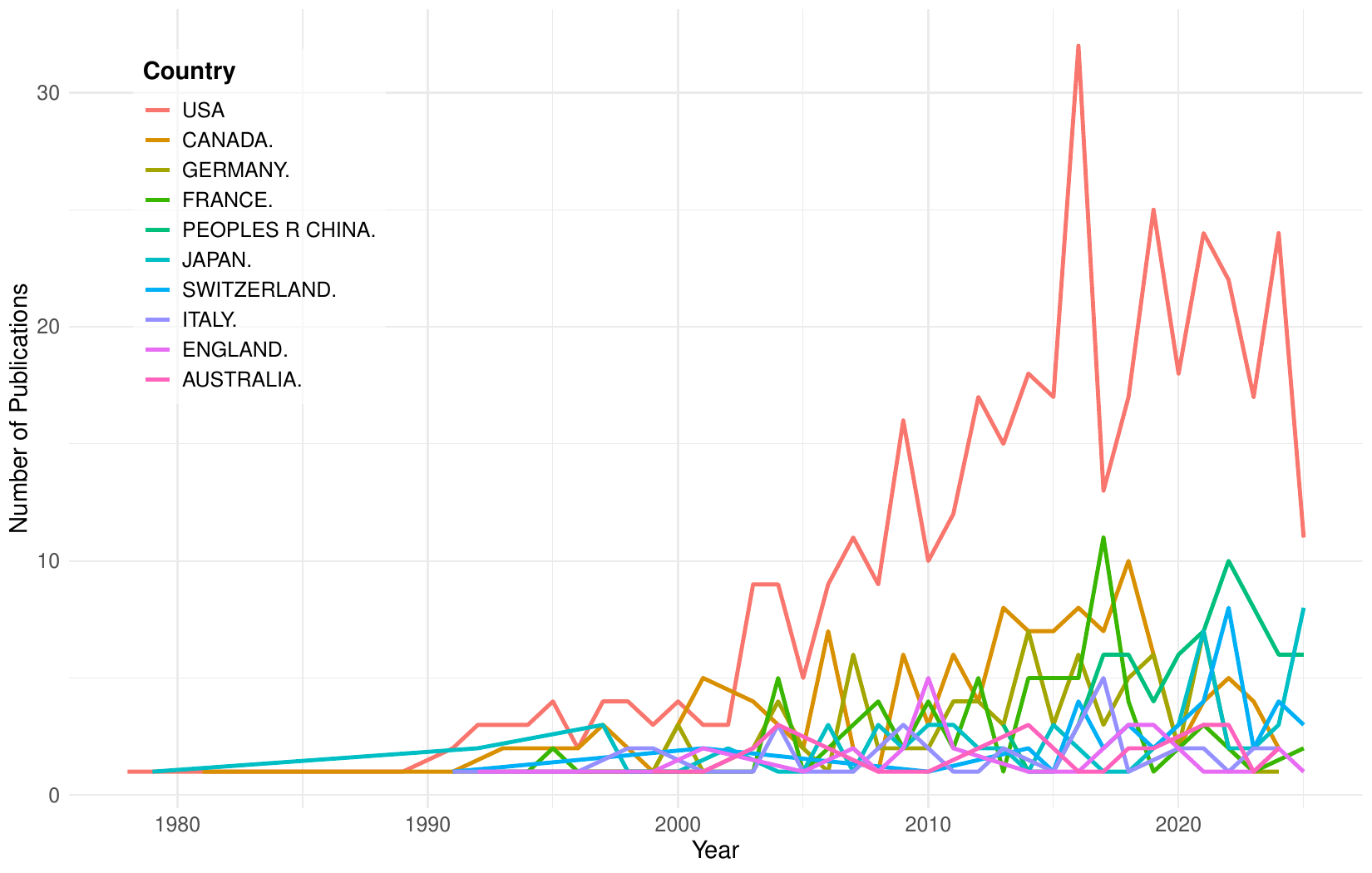}
    \caption{Publication trends over the years by country}
    \label{fig:country}
\end{figure}

\begin{figure}[h!]
    \centering
    \begin{subfigure}{0.32\textwidth}
        \includegraphics[width=\linewidth]{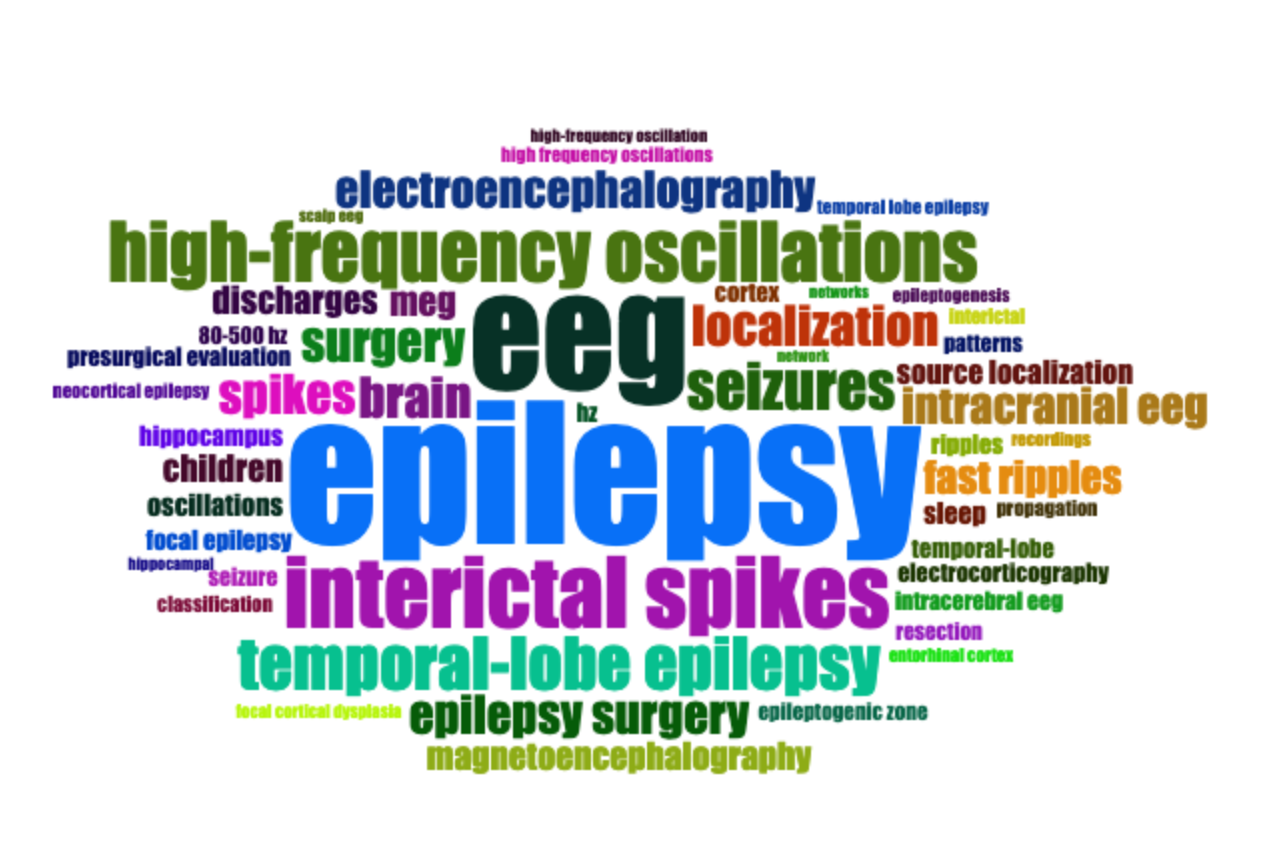}
        \caption{All-time} 
        \label{fig:wordclouds_a}
    \end{subfigure}
    \hfill
    \begin{subfigure}{0.32\textwidth}
        \includegraphics[width=\linewidth]{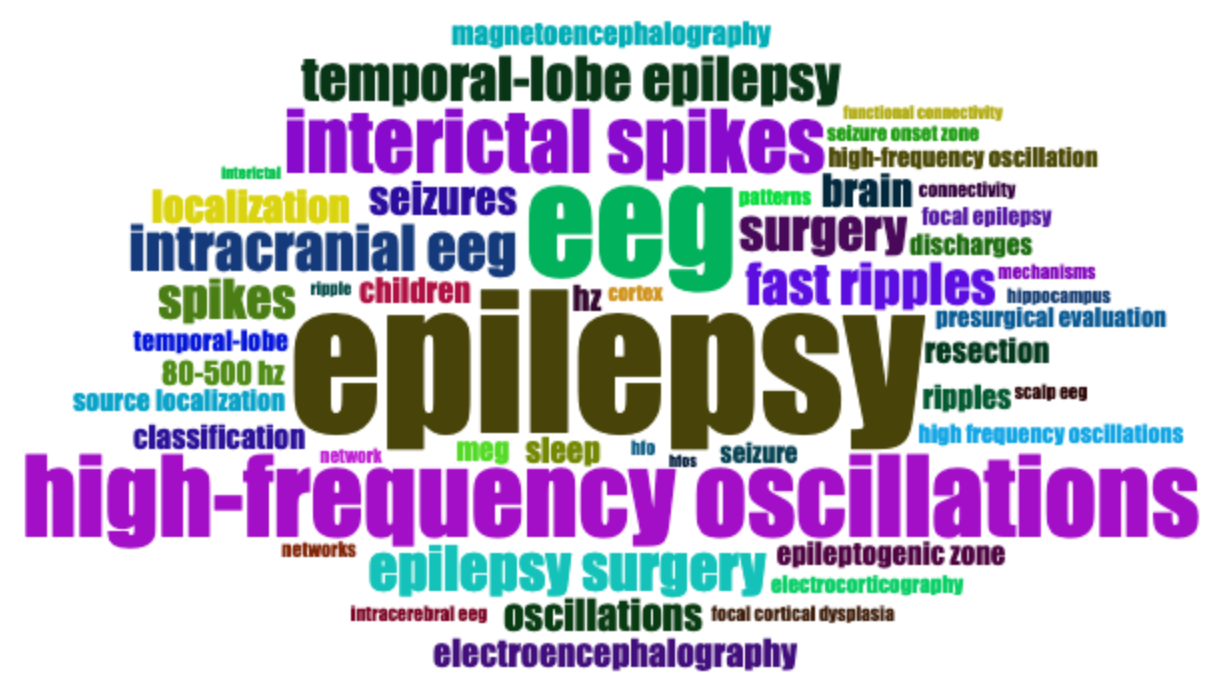}
        \caption{Past 10 years} 
        \label{fig:wordclouds_b}
    \end{subfigure}
    \hfill
    \begin{subfigure}{0.32\textwidth}
        \includegraphics[width=\linewidth]{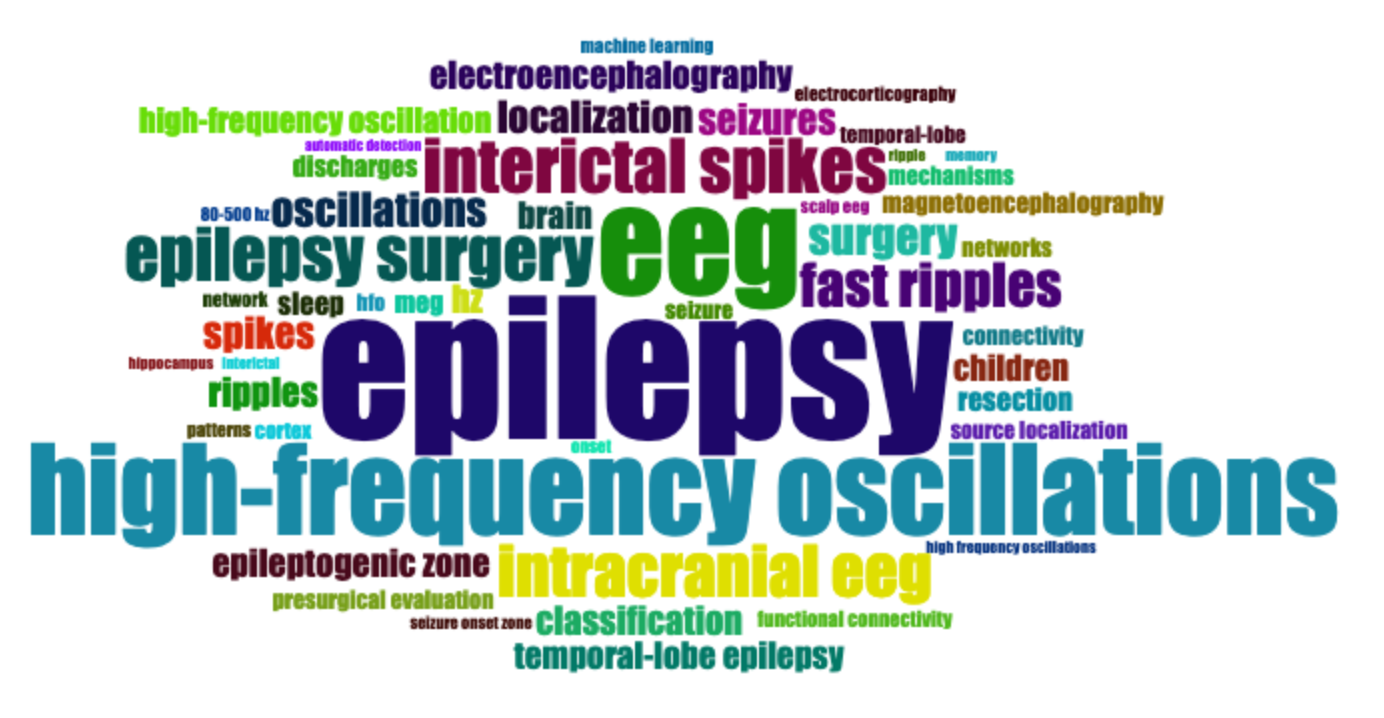}
        \caption{Past 5 years} 
        \label{fig:wordclouds_c}
    \end{subfigure}
    \caption{Word clouds of keywords from all-time (a), the past 10 years (b), and the past 5 years (c), illustrating the increasing prominence of ``HFO''.}
    \label{fig:wordclouds}
\end{figure}

\begin{figure}[h!]
    \centering
    \includegraphics[width=\textwidth]{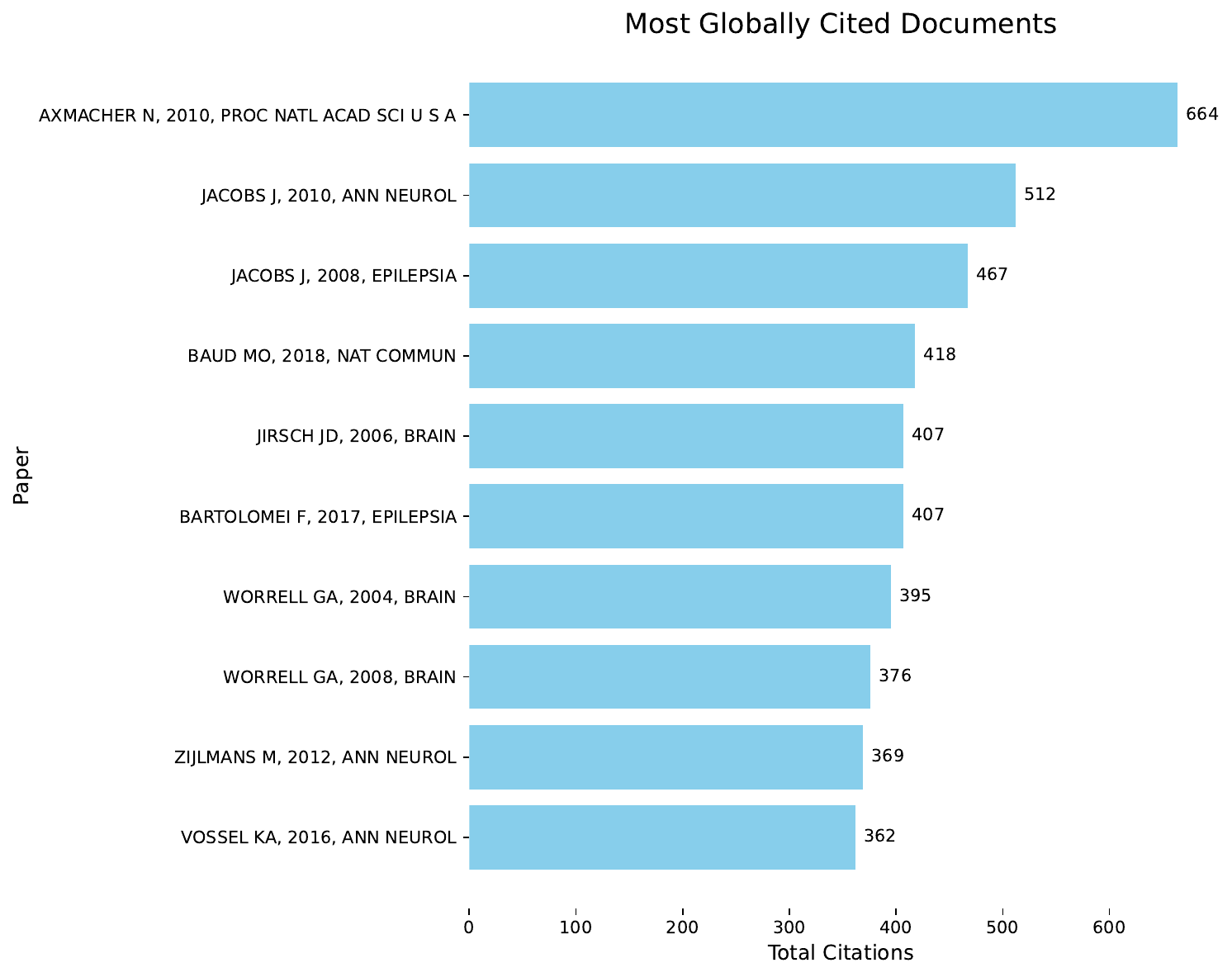}
    \caption{Top 10 most globally cited documents from the query.}
    \label{fig:cited}
\end{figure}

As illustrated in Figure \ref{fig:country}, research in this field is globally distributed and has experienced substantial growth in publication volume over the past decade. The United States has been the predominant contributor since the inception of this research area, a leadership role that has become even more pronounced during the recent acceleration in scholarly output. Word cloud analyses (Figure \ref{fig:wordclouds}) demonstrate a significant shift in focus over time. While ``epilepsy'' and ``spikes'' have been consistently prominent, the term ``HFO'' has grown dramatically in relevance, particularly in the last ten years, indicating its emergence as a central topic of investigation. Furthermore, an analysis of the most globally cited documents (Figure \ref{fig:cited}) highlights highly influential papers within this bibliometric scope, many of which focus on establishing the relationship between HFOs, epileptogenicity, and cognitive dysfunction.

\subsection{Scope and Structure of this Review}
This review provides a comprehensive analysis of publicly available EEG datasets relevant to the study of HFOs in AD and related neurodegenerative disorders. It aims to consolidate information and clarify nomenclature, and provide a detailed methodological synthesis. The following sections will:
\begin{itemize}
    \item Provide in-depth profiles of individual datasets, detailing their cohorts, acquisition parameters, and access information, with a focus on their suitability for HFO analysis.
    \item Discuss methodological heterogeneity of datasets and its implications for robust and reproducible HFO research.
    \item Conclusion and future work on the role of HFOs as a cross-disease biomarker.
\end{itemize}
\section{Publicly Available Datasets for HFO Analysis}

This section compares several key publicly available EEG datasets relevant to Alzheimer's Disease research. Each profile systematically outlines participant characteristics, data acquisition protocols, and access modalities. This structured presentation is designed to equip researchers with the specific information needed to select appropriate datasets for their scientific inquiries.

\begin{longtable}{@{}p{2.2cm} p{3.2cm} p{2.2cm} p{1.6cm} p{1.6cm} p{2cm}@{}}
\caption{Comparative Summary of Public EEG Datasets for AD and HFO Research}\\
\toprule
    \textbf{Dataset Name \& Repository} & 
    \textbf{Primary Citation} & 
    \textbf{Subjects} & 
    \textbf{Recording Type} & 
    \textbf{Sampling Freq. (Hz)} & 
    \textbf{HFO Compatible}\\
\midrule
\endfirsthead
\caption{(continued)}\\
\toprule
    \textbf{Dataset Name \& Repository} & 
    \textbf{Primary Citation} & 
    \textbf{Subjects} & 
    \textbf{Recording Type} & 
    \textbf{Sampling Freq. (Hz)} & 
    \textbf{HFO Compatible}\\
\midrule
\endhead
\bottomrule
\endfoot
\bottomrule
    \textbf{ADFTD} & 
    Miltiadous, A., et al. (2023) \cite{Miltiadous2023} &
    \vtop{
        \hbox{\strut \textbf{AD: 36}}
        \hbox{\strut FTD: 23}
        \hbox{\strut Control: 29}
    } &
    Scalp EEG &
    500 & 
    Yes (Ripples) \\
\addlinespace
    \textbf{PEARL-Neuro} & 
    Dzianok, P., \& Kublik, E. (2024) \cite{Dzianok2024} & 
    \textbf{At risk of dementia: 79}& 
    Scalp EEG (High-Density) & 
    1000 & 
    Yes \\
\addlinespace
    \textbf{BrainLat} & 
    Prado, P., et al. (2023) \cite{Prado2023} &
    \vtop{
        \hbox{\strut \textbf{AD: 278}}
        \hbox{\strut bvFTD: 163}
        \hbox{\strut PD: 57}
        \hbox{\strut MS: 32}
        \hbox{\strut Control: 250}
    } &
    Scalp EEG (High-Density) & 
    512 & 
    Yes (Ripples) \\
\addlinespace
    \textbf{Pineda et al.} & 
    Pineda, A. M., et al. (2020) \cite{Pineda2020} & 
    \vtop{
        \hbox{\strut \textbf{probable AD: 24}}
        \hbox{\strut Control: 24}
    } & 
    Scalp EEG & 
    128 & 
    No\\
\addlinespace
    \textbf{Vicchietti et al.} & 
    Vicchietti, M. L., et al. (2025) \cite{Vicchietti2025} &
    \vtop{
        \hbox{\strut \textbf{probable AD: 160}}
        \hbox{\strut Control: 24}
    } &  
    Scalp EEG & 
    128 & 
    No \\
\addlinespace
    \textbf{Escudero et al.} & 
    Escudero, J., et al. (2006) \cite{Escudero2006} & 
    \vtop{
        \hbox{\strut \textbf{AD: 11}}
        \hbox{\strut Control: 11}
    } & 
    Scalp EEG & 
    256 & 
    Yes (Ripples) \\
\bottomrule
\end{longtable}
\section{Controlled-Access Datasets for Neurodegenerative and Neurological Research}

Large-scale, longitudinal datasets are indispensable for identifying biomarkers and understanding the progression of neurological disorders. To the best of our knowledge, most comprehensive datasets are housed in controlled-access repositories to protect sensitive patient information. Access typically requires researchers to submit a formal application detailing their research plan and agreeing to data use policies.

A landmark example is the Alzheimer's Disease Neuroimaging Initiative (ADNI), a longitudinal study essential for researchers investigating the progression from healthy aging to Mild Cognitive Impairment (MCI) and Alzheimer's disease \cite{weiner2017alzheimer}. Its primary strength lies in its extensive collection of imaging data (MRI, PET) and biological markers (CSF, blood). Due to its comprehensive, longitudinal nature, ADNI remains a cornerstone for multimodal research. Access to all ADNI data is managed through the Laboratory of Neuro Imaging (LONI) Image \& Data Archive (IDA), which requires a formal online application.

Similarly, the Parkinson's Progression Marker Initiative (PPMI) is a major multi-site study focused on identifying biomarkers of Parkinson's disease (PD) progression through comprehensive, longitudinal data collection \cite{marek2018parkinson}. The initiative has built a rich repository containing extensive imaging, biological (e.g., biospecimens), and pathological data from PD patients, prodromal subjects, and healthy controls. This robust dataset is designed to support the validation of biomarkers that track the course of PD. Researchers can request access to this resource through the PPMI database after completing a registration process and signing a data use agreement.

For research focusing on magnetoencephalography (MEG), the BioFIND dataset is a vital, publicly available resource \cite{vaghari2021biofind}. This multi-site, multi-participant resting-state MEG collection was specifically curated to advance the study of dementia. It contains data from individuals with Mild Cognitive Impairment (MCI), subjective cognitive impairment, and healthy controls, providing the raw neuromagmetic data crucial for investigating alterations in brain oscillations and functional connectivity. Access is granted upon valid request, facilitating collaborative research into MEG-based biomarkers for dementia.

Finally, while focused on epilepsy, the clinical study titled Clinical Applications of High-frequency Oscillations (HFOs), led by Dr. Jing Xiang, has generated a methodologically significant dataset \cite{xiang2008clinical}. The high-resolution MEG and EEG data collected from pediatric patients are highly relevant for any research investigating pathological high-frequency brain activity \cite{xiang2021}. Researchers interested in utilizing this specialized dataset can typically gain access by submitting a formal request to the principal investigator.
\section{Navigating Methodological Heterogeneity: Challenges and Opportunities}

The diversity across the datasets, while offering a broad research landscape, also introduces significant methodological challenges. The variations in sampling frequency, preprocessing pipelines, and data accessibility are not trivial details; they are fundamental sources of analytic variance that can impact the reproducibility and generalizability of research findings. Understanding and navigating this heterogeneity is crucial for future research aiming to use or combine these public resources.

\subsection{The Impact of Sampling Frequency on Analysis}

The range in EEG sampling frequencies—128 Hz \cite{Pineda2020} to 500 Hz \cite{Miltiadous2023, Prado2023} to 1000 Hz \cite{Dzianok2024}—is one of the most critical dimensions of heterogeneity. The sampling frequency ($f_s$) of a digital signal determines the highest frequency that can be unambiguously represented, known as the Nyquist frequency ($f_s / 2$). Early investigations of very fast activity during seizures highlighted the need for wide bandwidths \cite{Allen1992}.

For the Pineda et al. dataset, with its 128 Hz sampling rate, the Nyquist frequency is 64 Hz. While this is adequate for analyzing the canonical EEG bands (delta, theta, alpha, beta), it makes the analysis of the gamma band (typically defined as 30-100 Hz) difficult, as frequencies above 64 Hz are aliased and cannot be distinguished from lower frequencies. In contrast, the 500 Hz sampling rate of the ADFTD dataset provides a comfortable margin for analyzing the full gamma spectrum and even the lower end of the ripple band (80-250 Hz), with a Nyquist frequency of 250 Hz. The 1000 Hz rate of the PEARL-Neuro dataset pushes this even further to 500 Hz, enabling the exploration of the full range of HFOs, including fast ripples (250-500 Hz).

This heterogeneity has two major consequences. First, a research question focused on high-frequency activity is immediately constrained to using datasets like ADFTD or PEARL-Neuro. Second, any attempt to perform meta-analysis or cross-dataset validation that includes datasets with different sampling rates requires a careful harmonization step. The most common approach is to downsample all datasets to the lowest common sampling rate. For instance, to combine the Pineda et al. and ADFTD datasets, the latter would need to be downsampled from 500 Hz to 128 Hz, a process that involves low-pass filtering to prevent aliasing, followed by decimation. This ensures that all data are processed on an equal footing, but it comes at the cost of discarding the high-frequency information present in the higher-quality recording.

\subsection{Preprocessing Pipelines and Source Localization}

EEG data is inherently noisy, contaminated by artifacts from eye movements, muscle activity, and environmental electrical interference. The preprocessing pipeline can have a profound impact on the final results. The ADFTD dataset provides an excellent case study in transparent, modern preprocessing, detailing steps such as band-pass filtering (0.5-45 Hz), re-referencing, and advanced artifact correction using both Artifact Subspace Reconstruction (ASR) and Independent Component Analysis (ICA) \cite{Miltiadous2023}. The development of automated spike and HFO detection algorithms, often based on wavelet transforms or other time-frequency methods, contributes to the development of standardized approaches for identifying these transient events. \cite{Tieng2016, McCall2012, LOPEZCUEVAS2013, Wang2022, Wang2024}.

The challenge arises because there is no single, universally accepted pipeline. Different researchers may choose different filter cutoffs, reference schemes, or artifact rejection methods. These choices are not neutral. An aggressive filter might remove neural signal along with noise, while a lenient one might leave residual artifacts that could be misinterpreted as pathological brain activity. Consequently, it is often difficult to directly compare the results of a machine learning classifier trained on data from one study with the results from another if the preprocessing pipelines differ. The choice of source localization technique, such as sLORETA, also introduces another layer of analytic variability, although such methods are crucial for inferring the cortical origins of scalp activity \cite{Eom2017}.

This challenge also presents an opportunity for methodological refinement. The availability of both raw and preprocessed data in BIDS-compliant datasets such as ADFTD enables structured evaluation of preprocessing strategies. Researchers can assess how different preprocessing choices affect diagnostic outcomes when applied to a shared raw dataset, clarifying the influence of early-stage decisions on analytical performance. Furthermore, the variability across datasets provides a practical framework for testing algorithmic robustness. A classification method that performs exclusively on a single, meticulously curated dataset lacks clinical utility. In contrast, an approach that achieves consistent and accurate results across datasets with differing technical specifications and preprocessing protocols demonstrates robustness and applicability to clinical practice.
\section{Data Accessibility and the Importance of FAIR Principles}

A final, pragmatic challenge in neurophysiological research lies in the accessibility and completeness of the data. While the ideal is open science, the reality is a landscape composed of both public and non-public, controlled-access datasets. This review identified several instances where data access was problematic. For older datasets like Escudero et al. (2006), no stable public access link could be found, rendering it effectively unavailable for new research and highlighting the problem of data decay.

This issue is partially addressed by modern data sharing practices, but it also underscores the critical role of non-public datasets. A significant portion of valuable clinical data, particularly longitudinal studies with sensitive patient information like the Alzheimer's Disease Neuroimaging Initiative (ADNI) or the Parkinson's Progression Marker Initiative (PPMI), cannot be made fully public. These non-public datasets are accessible only upon valid request, governed by strict data use agreements to protect patient privacy.

These challenges highlight a lack of standardization and emphasize the vital importance of adhering to FAIR (Findable, Accessible, Interoperable, and Reusable) principles for all datasets, whether public or controlled. Even for non-public data, FAIR principles apply:
\begin{itemize}
    \item The data should be \textbf{Findable} through public metadata and a stable DOI.
    \item It should be \textbf{Accessible} under clear, well-defined conditions.
    \item It must be \textbf{Interoperable} and \textbf{Reusable} by using community standards like the Brain Imaging Data Structure (BIDS) and providing comprehensive metadata.
\end{itemize}

Datasets published recently on platforms like OpenNeuro (e.g., PEARL-Neuro, ADFTD) exemplify this modern approach, with dedicated publications and rich metadata. Adhering to these standards ensures that all valuable scientific data—public or controlled—remains a lasting and reliable resource. Furthermore, the complementary roles of EEG and magnetoencephalography (MEG) in spike detection highlight the need for multimodal data sharing, as combining them improves localization accuracy \cite{Lin2003, Ochi2001, Bagic2015}.
\section{Future Directions}

The synthesis of existing research and datasets indicates a field of significant scientific promise, yet one that also faces substantial methodological and translational hurdles. Although the evidence supporting high frequency oscillations (HFOs) as a biomarker of network hyperexcitability is compelling, its transition from a research finding to a clinically actionable tool depends on a coordinated effort to address critical gaps. The following directions represent a roadmap for navigating these challenges and realizing the full potential of HFOs in both epilepsy and neurodegenerative disease.

\begin{enumerate}
    \item \textbf{Standardization of Acquisition and Analytical Pipelines.} A primary obstacle to the robust meta-analysis and validation of HFOs is the methodological heterogeneity across studies. Future work must prioritize the development of standardized protocols for data acquisition and analysis, particularly for scalp EEG. Establishing consensus on best practices for filtering, artifact rejection, and the statistical definition of pathological events is essential to mitigate analytic variance and ensure cross-study reproducibility. The dissemination of open-source, validated computational toolboxes will be a cornerstone of this effort.

    \item \textbf{Advancement of Generalizable and Interpretable AI Models.} The AI models developed to date demonstrate high performance on specific datasets, but their clinical utility hinges on broader applicability. The next generation of algorithms must be designed to enhance model generalizability across diverse patient populations and recording equipment. This requires training on larger, more heterogeneous data and developing techniques that are robust to inter-dataset variability. Concurrently, addressing the ``black box'' problem by improving model interpretability is paramount for gaining the trust of clinicians and elucidating the specific neurophysiological features that drive diagnostic classifications.

    \item \textbf{Development of Longitudinal, Multimodal, and Diverse Cohorts.} The preponderance of cross-sectional datasets limits the current understanding of HFO dynamics over time. There is a profound need for more longitudinal initiatives that track individuals over multiple years. Such cohorts are invaluable for transitioning from diagnostic (cross-sectional) to prognostic (longitudinal) biomarkers, enabling the modeling of disease progression and the prediction of cognitive decline. These future data collection efforts must also prioritize multimodal integration (EEG, imaging, fluid biomarkers) and greater population diversity to ensure that findings are equitably applicable across different genetic and environmental backgrounds.

    \item \textbf{Prospective Validation and Clinical Implementation.} The ultimate objective is the clinical implementation of HFOs as a validated biomarker. This requires moving beyond retrospective analyses to large-scale, prospective clinical trials. For Alzheimer's disease, such trials must validate the diagnostic and prognostic utility of scalp HFOs against established gold-standard biomarkers. For epilepsy, the focus will be on integrating real-time HFO detection into clinical monitoring workflows and exploring its utility in guiding therapeutic interventions, including closed-loop neuromodulation systems.
\end{enumerate}

Addressing these imperatives will require a multidisciplinary and collaborative approach, bridging clinical neurology, data science, and biomedical engineering. Successfully navigating these future directions will not only solidify the role of HFOs in the clinical armamentarium but also deepen our fundamental understanding of network hyperexcitability as a transdiagnostic feature of major neurological disorders.
\section{Conclusion}

This review has synoptically charted the evolution of High-Frequency Oscillations (HFOs) from a specific biomarker in epilepsy to a promising indicator of network hyperexcitability in neurodegenerative disorders such as Alzheimer's disease. The bibliometric analysis confirms an accelerating research trajectory, underscoring the growing recognition of HFOs' clinical significance. Our detailed examination of the data landscape reveals a maturing ecosystem of both public and controlled-access repositories, which, despite significant methodological heterogeneity, provides the foundation for advanced computational analysis.

The availability of these datasets has directly enabled the development of sophisticated AI applications capable of diagnosing disease, localizing pathology, and predicting clinical outcomes with increasing accuracy. However, the synthesis of current research also illuminates the path forward. The successful clinical translation of HFOs is contingent upon addressing the critical challenges of methodological standardization, the development of generalizable and interpretable AI models, and rigorous validation through large-scale, prospective longitudinal studies. By navigating these future directions, the research community is poised to establish HFOs as a validated, non-invasive, and clinically actionable biomarker, which would fundamentally advance the diagnostic and therapeutic paradigms for a range of neurological diseases.

\printbibliography

@ARTICLE{Zijlmans2012,
title = {A comparison between detectors of high frequency oscillations},
journal = {Clinical Neurophysiology},
volume = {123},
number = {1},
pages = {106-116},
year = {2012},
issn = {1388-2457},
doi = {https://doi.org/10.1016/j.clinph.2011.06.006},
author = {R. Zelmann and F. Mari and J. Jacobs and M. Zijlmans and F. Dubeau and J. Gotman},

}

@article{Bragin1999,
    Author = {Bragin, A. and Engel, Jr., J. and Wilson, C. L. and Fried, I. and Buzsaki, G.},
    Title = {High-frequency oscillations in human brain},
    Journal = {Hippocampus},
    Year = {1999},
    Volume = {9},
    Number = {2},
    Pages = {137-142},
    DOI = {10.1002/(SICI)1098-1063(1999)9:2<137::AID-HIPO3>3.0.CO;2-1}
}

@article{Staba2002,
    Author = {Staba, R. J. and Wilson, C. L. and Bragin, A. and Fried, I. and Engel, Jr., J.},
    Title = {Quantitative analysis of high-frequency oscillations (80-500 Hz) recorded in human epileptic hippocampus and entorhinal cortex},
    Journal = {Journal of Neurophysiology},
    Year = {2002},
    Volume = {88},
    Number = {4},
    Pages = {1743-1752},
    DOI = {10.1152/jn.2002.88.4.1743}
}

@article{Worrell2011,
    Author = {Worrell, Greg and Gotman, Jean},
    Title = {High-frequency oscillations and other electrophysiological biomarkers of epilepsy: clinical studies},
    Journal = {Biomarkers in Medicine},
    Year = {2011},
    Volume = {5},
    Number = {5},
    Pages = {557-566},
    DOI = {10.2217/BMM.11.74}
}

@article{Jacobs2011,
    Author = {Jacobs, Julia},
    Title = {High Frequency Oscillation ($>$80 Hz) as Markers of Epileptogenicity},
    Journal = {Klinische Neurophysiologie},
    Year = {2011},
    Volume = {42},
    Number = {1},
    Pages = {9-16},
    DOI = {10.1055/s-0031-1271787}
}

@article{Buzsaki2015,
    Author = {Buzsáki, György},
    Title = {Hippocampal sharp wave-ripple: A cognitive biomarker for navigation and memory},
    Journal = {Neuron},
    Year = {2015},
    Volume = {85},
    Number = {6},
    Pages = {1175-1189},
    DOI = {10.1002/hipo.22488}
}

@article{Gerstl2023,
    Author = {Gerstl, Jakob V. E. and Kiseleva, Alina and Imbach, Lukas and Sarnthein, Johannes and Fedele, Tommaso},
    Title = {High frequency oscillations in relation to interictal spikes in predicting postsurgical seizure freedom},
    Journal = {Scientific Reports},
    Year = {2023},
    Volume = {13},
    Number = {1},
    pages = {21313},
    DOI = {10.1038/s41598-023-48764-4}
}

@article{Frauscher2018,
    Author = {Frauscher, B. and von Ellenrieder, N. and Zelmann, R. and Rogers, C. and Nguyen, D. K. and Kahane, P. and Dubeau, F. and Gotman, J.},
    Title = {High-frequency oscillations in the normal human brain},
    Journal = {Annals of Neurology},
    Year = {2018},
    Volume = {84},
    Number = {3},
    Pages = {374-385},
    DOI = {10.1002/ana.25304}
}

@article{Maeda2025,
    Author = {Maeda, Keisuke and Hosoda, Nami and Fukumoto, Junichi and Kawai, Shun and Hayafuji, Mizuki and Tsuboi, Himari and Fujita, Shiho and Ichino, Naohiro and Osakabe, Keisuke and Sugimoto, Keiko and Ishihara, Naoko},
    Title = {Association of Scalp High-Frequency Oscillation Detection and Characteristics With Disease Activity in Pediatric Epilepsy},
    Journal = {Journal of Clinical Neurophysiology},
    Year = {2025},
    Volume = {42},
    Number = {1},
    Pages = {28-35},
    DOI = {10.1097/WNP.0000000000001052}
}

@article{deCurtis2009,
    Author = {de Curtis, Marco and Gnatkovsky, Valerio},
    Title = {Re-evaluating the mechanisms of focal ictogenesis: The role of low-voltage fast activity},
    Journal = {Epilepsia},
    Year = {2009},
    Volume = {50},
    Number = {12},
    Pages = {2514-2525},
    DOI = {10.1111/j.1528-1167.2009.02242.x}
}

@ARTICLE{Staley2006,
  title    = "Interictal spikes and epileptogenesis",
  author   = "Staley, Kevin J and Dudek, F Edward",
  journal  = "Epilepsy Curr",
  volume   =  6,
  number   =  6,
  pages    = "199--202",
  year     =  2006,
  address  = "United States",
  DOI = {10.1111/j.1535-7511.2006.00145.x}
}

@article{Guth2021,
    Author = {Guth, Tim A. and Kunz, Lukas and Brandt, Armin and Duempelmann, Matthias and Klotz, Kerstin A. and Reinacher, Peter C. and Schulze-Bonhage, Andreas and Jacobs, Julia and Schoenberger, Jan},
    Title = {Interictal spikes with and without high-frequency oscillation have different single-neuron correlates},
    Journal = {Brain},
    Year = {2021},
    Volume = {144},
    Number = {10},
    Pages = {3078-3088},
    DOI = {10.1093/brain/awab288}
}

@ARTICLE{Haegelen2013,
  title    = "High-frequency oscillations, extent of surgical resection, and
              surgical outcome in drug-resistant focal epilepsy",
  author   = "Haegelen, Claire and Perucca, Piero and Ch{\^a}tillon,
              Claude-Edouard and Andrade-Valen{\c c}a, Luciana and Zelmann,
              Rina and Jacobs, Julia and Collins, D Louis and Dubeau, Fran{\c
              c}ois and Olivier, Andr{\'e} and Gotman, Jean",
  journal  = "Epilepsia",
  volume   =  54,
  number   =  5,
  pages    = "848--857",
  year     =  2013,
  DOI = {10.1111/epi.12075}
}

@ARTICLE{Holler2015,
  title    = "High-frequency oscillations in epilepsy and surgical outcome. A
              meta-analysis",
  author   = "H{\"o}ller, Yvonne and Kutil, Raoul and Klaffenb{\"o}ck, Lukas
              and Thomschewski, Aljoscha and H{\"o}ller, Peter M and Bathke,
              Arne C and Jacobs, Julia and Taylor, Alexandra C and Nardone,
              Raffaele and Trinka, Eugen",journal  = "Front Hum Neurosci",
  volume   =  9,
  pages    = "574",
  year     =  2015,
  DOI= {10.3389/fnhum.2015.00574}
}

@article{vantKlooster2015,
author = {Maryse A. van 't Klooster  and Nicole E.C. van Klink  and Frans S.S. Leijten  and Rina Zelmann  and Tineke A. Gebbink  and Peter H. Gosselaar  and Kees P.J. Braun  and Geertjan J.M. Huiskamp  and Maeike Zijlmans },
title = {Residual fast ripples in the intraoperative corticogram predict epilepsy surgery outcome},
journal = {Neurology},
volume = {85},
number = {2},
pages = {120-128},
year = {2015},
doi = {10.1212/WNL.0000000000001727},
}

@article{Wu2010,
    Author = {Wu, J. Y. and Sutherling, W. W. and Koh, S. and Matsumoto, J. H. and Ason, G. T. and Salamon, N.},
    Title = {Removing interictal fast ripples on electrocorticography linked with seizure freedom in children},
    Journal = {Neurology},
    Year = {2010},
    Volume = {75},
    Number = {19},
    Pages = {1686-1694},
    DOI = {10.1212/WNL.0b013e3181fc27d0}
}

@article{Holmes2013,
    Author = {Holmes, Gregory L.},
    Title = {EEG abnormalities as a biomarker for cognitive comorbidities in pharmacoresistant epilepsy},
    Journal = {Epilepsia},
    Year = {2013},
    Volume = {54 Suppl 2},
    Pages = {60-62},
    DOI = {10.1111/epi.12186}
}

@article{Kleen2013,
    Author = {Kleen, J. K. and Scott, R. C. and Holmes, G. L. and Lenck-Santini, P. P.},
    Title = {Hippocampal interictal epileptiform activity disrupts cognition in humans},
    Journal = {Neurology},
    Year = {2013},
    Volume = {81},
    Number = {1},
    Pages = {18-26},
    DOI = {10.1212/WNL.0b013e318297ee50}
}

@article{Vossel2016,
    Author = {Vossel, Keith A. and Ranasinghe, Kamalini G. and Beagle, Alexander J. and Mizuiri, Danielle and Honma, Susanne M. and Dowling, Anne F. and Darwish, Sonja M. and Van Berlo, Victoria and Barnes, Deborah E. and Mantle, Mary and Karydas, Anna M. and Coppola, Giovanni and Roberson, Erik D. and Miller, Bruce L. and Garcia, Paul A. and Kirsch, Heidi E. and Mucke, Lennart and Nagarajan, Srikantan S.},
    Title = {Incidence and Impact of Subclinical Epileptiform Activity in Alzheimer's Disease},
    Journal = {Annals of Neurology},
    Year = {2016},
    Volume = {80},
    Number = {6},
    Pages = {858-870},
    DOI = {10.1002/ana.24794}
}

@article{Palop2007,
    author={Palop, Jorge J.
    and Chin, Jeannie
    and Roberson, Erik D.
    and Wang, Jun
    and Thwin, Myo T.
    and Bien-Ly, Nga
    and Yoo, Jong
    and Ho, Kaitlyn O.
    and Yu, Gui-Qiu
    and Kreitzer, Anatol
    and Finkbeiner, Steven
    and Noebels, Jeffrey L.
    and Mucke, Lennart},
    title={Aberrant Excitatory Neuronal Activity and Compensatory Remodeling of Inhibitory Hippocampal Circuits in Mouse Models of Alzheimer's Disease},
    journal={Neuron},
    year={2007},
    month={Sep},
    day={06},
    publisher={Elsevier},
    volume={55},
    number={5},
    pages={697-711},
    issn={0896-6273},
    doi={10.1016/j.neuron.2007.07.025},
}

@article{Tamilia2021,
    Author = {Tamilia, E. and Toth, E. and Lane, E. M. and Vaudry, D. and Huneau, C. and Guedj, E. and Gotman, J. and Dubeau, F. and Kovac, S. and Bernard, G. and Carmant, L. and Kobayashi, E.},
    Title = {Noninvasive Mapping of Ripple Onset Predicts Outcome in Epilepsy Surgery},
    Journal = {Annals of Neurology},
    Year = {2021},
    Volume = {89},
    Number = {5},
    Pages = {911-923},
    DOI = {10.1002/ana.26066}
}

@article{Lisgaras2023,
    Author = {Lisgaras, Christos Panagiotis and Scharfman, Helen E.},
    Title = {High-frequency oscillations (250-500 Hz) in animal models of Alzheimer's disease and two animal models of epilepsy},
    Journal = {Epilepsia},
    Year = {2023},
    Volume = {64},
    Number = {1},
    Pages = {231-246},
    DOI = {10.1111/epi.17462}
}

@article{Verret2012,
    Author = {Verret, L. and Mann, E. O. and Hang, G. B. and Barth, A. M. I. and Cobos, I. and Ho, K. and Devidze, N. and Masliah, E. and Kreitzer, A. C. and Mody, I. and Mucke, L. and Palop, J. J.},
    Title = {Inhibitory Interneuron Deficit Links Altered Network Activity and Cognitive Dysfunction in a Mouse Model of Alzheimer's Disease},
    Journal = {Cell},
    Year = {2012},
    Volume = {149},
    Number = {3},
    Pages = {708-721},
    DOI = {10.1016/j.cell.2012.02.046}
}

@article{Miltiadous2023,
  author    = {Miltiadous, Andreas and Tzimourta, Katerina D. and Afrantou, Theodora and Ioannidis, Panagiotis and Grigoriadis, Nikolaos and Tsalikakis, Dimitrios G. and Angelidis, Pantelis and Tsipouras, Markos G. and Glavas, Evripidis and Giannakeas, Nikolaos and Tzallas, Alexandros T.},
  title     = {A Dataset of Scalp {EEG} Recordings of Alzheimer's Disease, Frontotemporal Dementia and Healthy Subjects from Routine {EEG}},
  journal   = {Data},
  volume    = {8},
  number    = {6},
  pages     = {95},
  year      = {2023},
  doi       = {10.3390/data8060095}
}

@article{Dzianok2024,
  author    = {Dzianok, Patrycja and Kublik, Ewa},
  title     = {{PEARL-Neuro} Database: {EEG}, {fMRI}, health and lifestyle data of middle-aged people at risk of dementia},
  journal   = {Scientific Data},
  volume    = {11},
  number    = {1},
  pages     = {276},
  year      = {2024},
  doi       = {10.1038/s41597-024-03106-5}
}

@article{Prado2023,
  author    = {Prado, Pavel and Medel, Vicente and Gonzalez-Gomez, Raul and Sainz-Ballesteros, Agustín and Vidal, Victor and Santamaría-García, Hernando and Moguilner, Sebastian and Mejia, Jhony and Slachevsky, Andrea and Behrens, Maria Isabel and Aguillon, David and Lopera, Francisco and Parra, Mario A. and Matallana, Diana and Maito, Marcelo Adrián and Garcia, Adolfo M. and Custodio, Nilton and Ávila Funes, Alberto and Piña-Escudero, Stefanie and Birba, Agustina and Fittipaldi, Sol and Legaz, Agustina and Ibañez, Agustín},
  title     = {The {BrainLat} project, a multimodal neuroimaging dataset of neurodegeneration from underrepresented backgrounds},
  journal   = {Scientific Data},
  volume    = {10},
  number    = {1},
  pages     = {889},
  year      = {2023},
  doi       = {10.1038/s41597-023-02806-8}
}

@article{Pineda2020,
  author    = {Pineda, Aruane M. and Ramos, Fernando M. and Betting, Luiz Eduardo and Campanharo, Andriana S. L. O.},
  title     = {Quantile graphs for {EEG}-based diagnosis of Alzheimer's disease},
  journal   = {PLoS One},
  volume    = {15},
  number    = {6},
  pages     = {e0231169},
  year      = {2020},
  doi       = {10.1371/journal.pone.0231169}
}

@misc{Vicchietti2025,
  author    = {Vicchietti, Mário L. and Ramos, Fernando M. and Betting, Luiz E. and Campanharo, Andriana S. L. O.},
  title     = {Data from: Computational methods of {EEG} signals analysis for Alzheimer's disease classification},
  year      = {2020},
  publisher = {OSF},
  doi       = {10.17605/OSF.IO/2V5MD}
}

@article{Escudero2006,
  author    = {Escudero, J. and Abásolo, D. and Hornero, R. and Espino, P. and López, M.},
  title     = {Analysis of electroencephalograms in Alzheimer's disease patients with multiscale entropy},
  journal   = {Physiological Measurement},
  volume    = {27},
  number    = {11},
  pages     = {1091-1106},
  year      = {2006},
  doi       = {10.1088/0967-3334/27/11/004}
}

@article{weiner2017alzheimer,
  title={The Alzheimer's disease neuroimaging initiative 3: Continued implementation of new modalities and clinical trial partnership},
  author={Weiner, Michael W and Veitch, Dallas P and Aisen, Paul S and Beckett, Laurel A and Cairns, Nigel J and Green, Robert C and Harvey, Danielle and Jack, Clifford R and Jagust, William and Luthman, Johan and others},
  journal={Alzheimer's \& Dementia},
  volume={13},
  number={5},
  pages={561--571},
  year={2017},
  doi = {10.1016/j.jalz.2016.10.006}
}

@article{marek2018parkinson,
  title={The Parkinson's Progression Markers Initiative (PPMI) - establishing a PD biomarker cohort},
  author={Marek, Kenneth and Chowdhury, Sohini and Siderowf, Andrew and Lasch, Shirley and Coffey, Christopher S and Caspell-Garcia, Chelsea and Simuni, Tanya and Jennings, Danna and Tanner, Caroline M and Kieburtz, Karl and others},
  journal={Annals of Clinical and Translational Neurology},
  volume={5},
  number={12},
  pages={1460--1477},
  year={2018},
  doi={10.1002/acn3.644},
}

@article{vaghari2021biofind,
  title={A multi-site, multi-participant magnetoencephalography resting-state dataset to study dementia: The BioFIND dataset},
  author={Vaghari, Delshad and Vakorin, Vasiliy A and versatile, bio-medical informatics and technology development},
  journal={NeuroImage},
  volume={244},
  pages={118598},
  year={2021},
  doi={10.1016/j.neuroimage.2022.119344}
}

@misc{xiang2008clinical,
  author={Xiang, Jing},
  title={Clinical Applications of High-frequency Oscillations (HFOs)},
  year={2008--2019},
  howpublished={ClinicalTrials.gov Identifier: NCT00600717},
  note={Accessed on August 24, 2025},
  url={https://clinicaltrials.gov/study/NCT00600717}
}

@article{xiang2021,
    author={Xiang, Jing
    and Maue, Ellen
    and Tong, Han
    and Mangano, Francesco T.
    and Greiner, Hansel
    and Tenney, Jeffrey},
    title={Neuromagnetic high frequency spikes are a new and noninvasive biomarker for localization of epileptogenic zones},
    journal={Seizure - European Journal of Epilepsy},
    year={2021},
    month={Jul},
    day={01},
    publisher={Elsevier},
    volume={89},
    pages={30-37},
    issn={1059-1311},
    doi={10.1016/j.seizure.2021.04.024},
}

@article{Allen1992,
    Author = {Allen, P. J. and Fish, D. R. and Smith, S. J.},
    Title = {Very high-frequency rhythmic activity during seizures in humans},
    Journal = {Electroencephalography and Clinical Neurophysiology},
    Year = {1992},
    Volume = {82},
    Number = {2},
    Pages = {155-159},
    DOI = {10.1016/0013-4694(92)90160-j}
}

@article{Tieng2016,
    Author = {Tieng, Quang M. and Kharatishvili, Irina and Chen, Min and Reutens, David C.},
    Title = {Mouse EEG spike detection based on the adapted continuous wavelet transform},
    Journal = {Journal of Neural Engineering},
    Year = {2016},
    Volume = {13},
    Number = {2},
    pages = {026018},
    DOI = {10.1088/1741-2560/13/2/026018}
}

@inproceedings{McCall2012,
    Author = {McCall, Paul and Cabrerizo, Mercedes and Adjouadi, Malek},
    Title = {Spatial and Temporal Analysis of Interictal Activity in the Epileptic Brain},
    Booktitle = {2012 IEEE Signal Processing in Medicine and Biology Symposium (SPMB)},
    Year = {2012},
    DOI = {10.1109/SPMB.2012.6469459}
}

@article{LOPEZCUEVAS2013,
title = {An algorithm for on-line detection of high frequency oscillations related to epilepsy},
journal = {Computer Methods and Programs in Biomedicine},
volume = {110},
number = {3},
pages = {354-360},
year = {2013},
issn = {0169-2607},
doi = {https://doi.org/10.1016/j.cmpb.2013.01.014},
author = {Armando López-Cuevas and Bernardino Castillo-Toledo and Laura Medina-Ceja and Consuelo Ventura-Mejía and Kenia Pardo-Peña},
}

@ARTICLE{Wang2022,
  title    = "Diagnostic value of high-frequency oscillations for the
              epileptogenic zone: A systematic review and meta-analysis",
  author   = "Wang, Yangshuo and Xu, Jinshan and Liu, Tinghong and Chen, Feng
              and Chen, Shuai and Yuan, Liu and Zhai, Feng and Liang, Shuli",
  journal  = "Seizure",
  volume   =  99,
  pages    = "82--90",
  month    =  may,
  year     =  2022,
  DOI = {10.1016/j.seizure.2022.05.003}
}

@ARTICLE{Wang2024,
  title    = "Prognostic Value of Complete Resection of the {High-Frequency}
              Oscillation Area in Intracranial {EEG}: A Systematic Review and
              {Meta-Analysis}",
  author   = "Wang, Ziyi and Guo, Jiaojiao and van 't Klooster, Maryse and
              Hoogteijling, Sem and Jacobs, Julia and Zijlmans, Maeike",
  journal  = "Neurology",
  volume   =  102,
  number   =  9,
  pages    = "e209216",
  month    =  apr,
  year     =  2024,
  DOI = {10.1111/epi.12213}
}

@article{Eom2017,
    Author = {Eom, Tae-Hoon and Shin, Jung-Hyun and Kim, Young-Hoon and Chung, Seung-Yun and Lee, In-Goo and Kim, Jung-Min},
    Title = {Distributed source localization of interictal spikes in benign childhood epilepsy with centrotemporal spikes: A standardized low-resolution brain electromagnetic tomography (sLORETA) study},
    Journal = {Journal of Clinical Neuroscience},
    Year = {2017},
    Volume = {38},
    Pages = {49-54},
    DOI = {10.1016/j.jocn.2016.12.047}
}

@article{Lin2003,
    Author = {Lin, Y. Y. and Shih, Y. H. and Hsieh, J. C. and Yu, H. Y. and Yiu, C. H. and Wong, T. T. and Yeh, T. C. and Kwan, S. Y. and Ho, L. T. and Yen, D. J. and Wu, Z. A. and Chang, M. S.},
    Title = {Magnetoencephalographic yield of interictal spikes in temporal lobe epilepsy. Comparison with scalp EEG recordings},
    Journal = {NeuroImage},
    Year = {2003},
    Volume = {19},
    Number = {3},
    Pages = {1115-1126},
    DOI = {10.1016/s1053-8119(03)00181-2}
}

@article{Ochi2001,
    Author = {Ochi, A. and Otsubo, H. and Sharma, R. and Hunjan, A. and Rutka, J. T. and Chuang, S. H. and Kamijo, K. and Yamazaki, T. and Quint, P. and Kurelowech, L. and Sobel, D. F. and Aung, M. H. and Snead, O. C.},
    Title = {Comparison of electroencephalographic dipoles of interictal spikes from prolonged scalp video-electroencephalography and magnetoencephalographic dipoles from short-term recording in children with extratemporal lobe epilepsy},
    Journal = {Journal of Child Neurology},
    Year = {2001},
    Volume = {16},
    Number = {9},
    Pages = {661-667},
    DOI = {10.1177/088307380101600907}
}

@article{Bagic2015,
    Author = {Bagic, Anto and Ebersole, John S.},
    Title = {Does MEG/MSI dipole variability mean unreliability?},
    Journal = {Clinical Neurophysiology},
    Year = {2015},
    Volume = {126},
    Number = {1},
    Pages = {209-211},
    DOI = {10.1016/j.clinph.2014.01.037}
}

\end{document}